\documentclass[amsmath, amssymb, preprintnumbers, showpacs, showkeys, aps,prb,superscriptaddress,twocolumn]{revtex4-1}
\usepackage{graphicx}
\usepackage{braket}
\usepackage{ulem}   
\usepackage{dsfont}
\usepackage{amsthm,amsmath,amsfonts,amssymb,verbatim,color}
\usepackage{bbold}
\usepackage{graphicx}
\usepackage[T1]{fontenc}
\usepackage[colorlinks=true,citecolor=blue,linkcolor=blue,urlcolor=blue]{hyperref}
\newcommand{\bs}[1]{{\boldsymbol{#1}}}

\newcommand{\pf}{\mathrm{Pf}}
\newcommand{\sgn}{\mathop{\mathrm{sgn}}}

\begin{document}
\preprint{}

\title{A topological classification of molecules and chemical reactions with a perplectic structure}

\author{Lukas Muechler}
\affiliation{Center for Computational Quantum Physics,The Flatiron Institute, New York, New York, 10010, USA}
\begin{abstract}	
In this paper, a topological classification of molecules and their chemical reactions is proposed on a single particle level .
We consider zero-dimensional electronic Hamiltonians in a real-space tight-binding basis with spinless time-reversal symmetry and an additional spatial reflection symmetry. The symmetry gives rise to a perplectic structure and suggests a $\mathbb{Z}_2$ invariant in form of a pfaffian, which can be captured by an entanglement cut. 
We apply our findings to a class of chemical reactions studied by Woodward and Hoffmann, where a reflection symmetry is preserved along a one-dimensional reaction path and argue that the topological classification should contribute to the rate constants of these reactions. More concretely, we find that a reaction takes place experimentally whenever the reactants and products can be adiabatically deformed into each other, while reactions that require a change of topological invariants have not been observed experimentally.
\end{abstract}
\date\today
\maketitle

\section{Introduction} 
\label{sec:intro}
\begin{figure*}[t]
 \includegraphics[width=1.6\columnwidth]{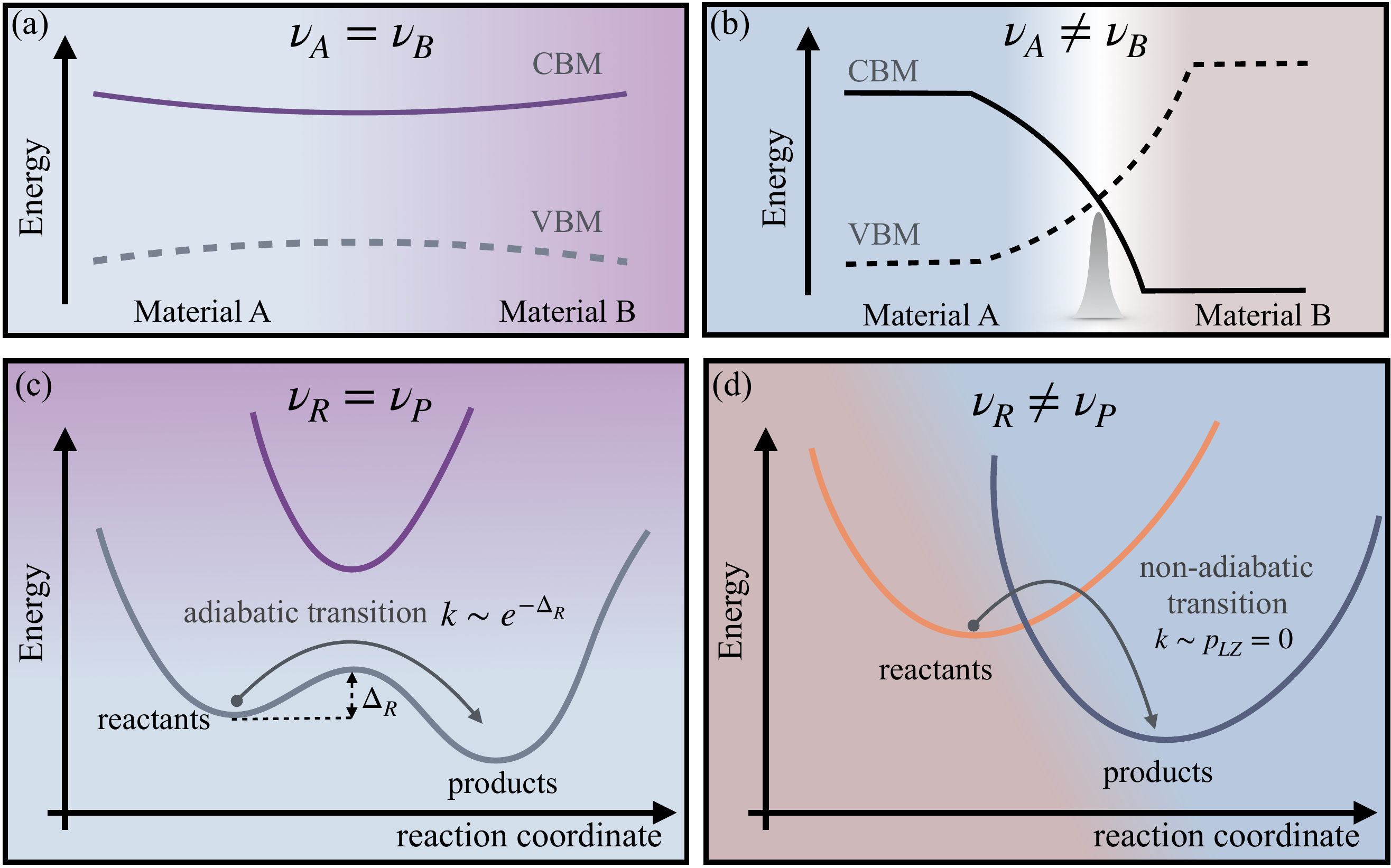}
\caption{
(a) An interface of two materials with the same topological invariant. While the valence band maxima (VBM) and conduction band maxima (CBM) change due to interface effects, the gap does not close at the interface
(b) An interface of two materials with different topological invariants. The spectral gap has to close at the interface and topologically protected surface states emerge.
(c) A chemical reaction in which the reactants and products have the same topological invariant. The spectral gap does not close along the reaction path and the reaction proceeds adiabatically and the reaction rate $k$ is determined by the activation barrier $\Delta_R$.
(d) A chemical reaction in which the reactants and products have a different topological invariant. Here, the spectral gap closes along the reaction path. The reaction rate is proportional to the probability $p_{RP}$ to transition between the GS of the reactants $\ket{R_0}$ to the ground state of the products $\ket{P_0}$, which can be estimated from Landau-Zener theory. 
}
\label{fig_1}
\end{figure*}
The advancements of the topological description of non-interacting crystalline matter within the last decade have revolutionized the field of condensed matter physics. \cite{TopoClass,ColloqBandTheory,yan2017topological,armitage2017weyl}
Topological considerations first predicted the existence of unremoveable exotic surface or edge states and have lead to the discovery of many new phases of matter in crystals \cite{QSH1,QSH2}, e.g. exotic fermionic states without analogues in high-energy physics.\cite{bradlyn2016beyond}
A state of matter is topologically non-trivial, if the ground state of the system cannot continuously be deformed into the atomic limit without gap-closing. 
Spatial and non-spatial symmetries determine the way this atomic limit is approached and give rise to a large variety of different non-interacting and interacting topological phases. \cite{kruthoff2016topological,bradlyn2017topological,po2017symmetry,watanabe2017structure}
These methods have recently found to be relevant for other areas of physics, e.g. mechanical systems, electric circuits and even weather phenomena \cite{kane2014topological,TopoSound,lee2017topolectrical,TopoWeather,cayssol2013floquet}\\
In like manner, it is known that topological effects play an important role in molecular systems and their chemical reactions. 
For example, the geometric phase acquired by moving around a conical intersection has been shown to strongly influence the reaction rate of simple chemical reactions due to interference of different reaction paths.~\cite{geophase1,h2h,butler1998chemical}
Further, it has been proposed that the surface states of Weyl semimetals or topological insulators could influence the outcome of chemical reactions as catalysts.~\cite{rajamathi2017weyl,rajamathi2017photochemical,politano2018toward,qu2018mbe} \\
Chemical reactions are rare events of the quantum dynamics on the Born-Oppenheimer (BO) surface and generally are complicated dynamical problems.
Despite empirical rules that provide strong guidelines, it is not fully understood why certain chemical reactions work the way they do, e.g. the Woodward-Hoffman rules (WHR).~\cite{JorgChem} 
In this paper we approach this problem from a topological point of view and propose a topological classification for molecules and their reactions that are described by these rules. The way this classification manifests itself in molecules is necessarily different from the solid state, since molecules are finite sized objects:
When two crystals with different topological invariants are brought in contact, the spectral gap has to close at the interface, since a topological invariant can only change at a gap closing point~\footnote{We only consider non-interacting systems in this paper. If interactions are included, invariants can change without gap closing}. At the interface, one therefore finds topologically required gapless states [Fig.~\ref{fig_1} (b)]. In molecules, such states will not appear due to their zero-dimensional (0D) nature; the interface states can generically be gapped out or are not well defined. \\
Instead, we here propose that a topological classification of molecules can manifest itself in their chemical reactions. 
We study chemical reactions of a set of \textit{reactant} molecules $R$ that transform into a set of \textit{product} molecules $P$~($R \rightarrow P$) and describe the transformation of the ground state (GS) of the reactant Hamiltonian $H_R$ into the GS of the Hamiltonian $H_P$ describing the products. A common way to model this process is to define a reaction path via a reaction coordinate $\tau$. This allows us to describe a chemical reaction as a continuous deformation of a reaction Hamiltonian $\mathcal{H}(\tau) = f(\tau) H_R + g(\tau) H_p$, with $f(0) = g(1) = 1$ and $f(1) = g(0) = 0$, which one can classify topologically. \\
This approach allows us to distinguish between two different cases as displayed in Fig.~\ref{fig_1}(c)\&(d). In the first case, reactants and products posses the same topological invariant. By definition, the GS of $H_R$ can be smoothly, i.e. adiabatically, deformed into the GS of $H_P$; concordantly the GS of $\mathcal{H}(\tau)$ is separated from the excited states by a gap for all $\tau$.
In the second case, the reactants and products differ in their topological invariants; the gap of $\mathcal{H}(\tau)$ has to close along the reaction path and the GS of reactants and products cannot be adiabatically transformed into each other, i.e. the reaction has to proceed in a non-adiabatic fashion. \\
The quantum mechanical observable associated with chemical reactions is the reaction rate. For adiabatic reactions, i.e. the first case, the Born-Oppenheimer approximation is usually valid and the reaction rate can be calculated by solely focusing on the GS of $\mathcal{H}(\tau)$. The rate is determined by the energy barrier of the reaction, which is determined by the energy of a transition state [Fig.~\ref{fig_1}(c)]. In the second case, transition state theory is not valid, as the higher energy states of $\mathcal{H}(\tau)$ cannot be ignored due to the gap closing point, as e.g. in the non-adiabatic regime of Marcus Theory.~\cite{marcus,blumberger2015recent} In a simplified picture, the reaction rate can be approximated from Landau Zener theory close to the gap closing point. For example, the rate can be computed as $k = \int dE \ p(E) e^{-\beta H}$, where $\beta$ is the inverse temperature, $H$ is the Hamiltonian in the micro-canonical ensemble and $p(E)$ is the probability to jump from the left surface to the right surface for a given energy $E$ [Fig.~\ref{fig_1}(d)]. 
Close to the crossing point, one can approximate this probability with the Landau-Zener probability $p(E) \simeq p_{LZ} = 1 - e^{- \xi \Delta^2}$, where $\Delta$ is the gap between the two potential energy surfaces and $\xi$ is a constant that depends on the details of the Hamiltonian. 
One therefore finds a vanishing reaction rate $k$ in case of a crossing ($\Delta = 0$), while it is exponentially small in the presence of a small gap. 
The physical picture is the following: instead of ending up in the GS of $H_P$, the final state of the reaction will be a linear combination of excited states and therefore the rate for the reaction, vanishes in this case.


\section{Outline}
The paper is structured as follows: We begin with the topological classification by introducing a simple toy-model in Sec.~\ref{sec:introex}, where we summarize and discuss the main results without derivation.
In the following section, Sec.~\ref{sec:bisymmetric}, we review the theory of bisymmetric matrices and derive the most general form of a spinless, time-reversal symmetric Hamiltonian of even matrix dimension with a reflection symmetry. In addition, we derive a general expression for the $\mathbb{Z}_2$topological invariant that has been introduced in the first section.\\
In Sec.~\ref{sec:chem}, we discuss the implications of the topological classification on the theory of chemical reactions, in particular the case of the Woodward-Hoffmann rules that are described by our models.

\section{A simple example} \label{sec:introex}
Before discussing the most general case, we start with a simple example of a $4\times4$ Hamiltonian $\mathcal{H}(t)$ at half filling, which depends on a single tuning parameter $t$ that takes on the role of the reaction coordinate while the other parameters $a,h,g$ remain constant. The Hamiltonian models the reaction of two ethylene molecules that approach each other along a reflection symmetric reaction pathway, which is modeled by tuning $t$ (see Sec.~\ref{sec:chem} for a more detailed discussion).
The Hamiltonian in the site basis is given as 
\begin{equation}
	\mathcal{H}(t) = 
	\left(
\begin{array}{cccc}
 g - \mu & t & h & a \\
 t & -g - \mu& a & h \\
 h & a & -g - \mu& t \\
 a & h & t & g - \mu\\
\end{array}
\right),
\end{equation}
where $\mu$ is the chemical potential, which will be used to define a suitable reference energy.
The Hamiltonian is time-reversal symmetric with $\mathcal{T} = \mathcal{K}$, where $\mathcal{K}$ is the anti-unitary complex conjugation operator. $\mathcal{H}(t)$ possesses a reflection symmetry $J$,
\begin{equation}
	J = 
\left(
\begin{array}{cccc}
 0 & 0 & 0 & 1 \\
 0 & 0 & 1 & 0 \\
 0 & 1 & 0 & 0 \\
 1 & 0 & 0 & 0 \\
\end{array}
\right)
\end{equation}
such that $[\mathcal{H}(t),J] = 0$ for all $t$. Note that the tuning-parameter $t$ is not affected by any symmetry.
Since $[\mathcal{H}(t),J] = 0$, there exists a orthogonal matrix $K$ which block-diagonalizes both $J$ and $\mathcal{H}(t)$,
\begin{equation}
	K = \frac{1}{\sqrt{2}} 
\left(
\begin{array}{cccc}
 -1 & 0 & 0 & 1 \\
 0 & -1 & 1 & 0 \\
 1 & 0 & 0 & 1 \\
 0 & 1 & 1 & 0 \\
\end{array}
\right).
\end{equation}
We arrive at
\begin{equation}
	\mathcal{H}_B(t) = K \mathcal{H}(t) K^T = 
	\left(
\begin{array}{cc}
H_{-}(t) & 0 \\ 
0 & H_{+}(t)  \\ 
\end{array}
\right),
\end{equation}
where the blocks corresponding to the $\pm$ eigenspaces of $J$ are obtained as 
$\mathcal{H}_{\pm} = \left(
\begin{array}{cc}
 \pm a+g-\mu  & t\pm h \\
 t\pm h & \pm a-g-\mu  \\
\end{array}
\right).$

\begin{figure*}[t]
 \includegraphics[width=1.4\columnwidth]{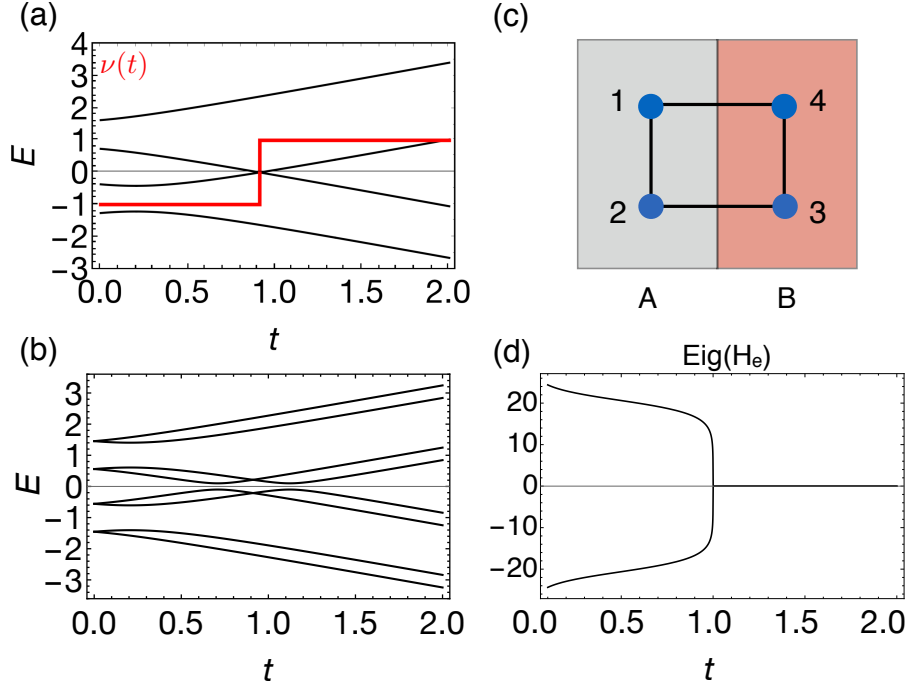}
\caption{
(a) Eigenvalues of $\mathcal{H}(t)$ and topological invariant $\nu(t)$ as function of $t$ for $ a = 1$, $g = 0.4$ and $h = 0.2$.
(b) Eigenvalues of the doubled Hamiltonian $\mathcal{H}_D(t)$ with $M = 0.1$
(c) Real space picture of entanglement cut and definition of subsystem $A$ and $B$.
(d) Entanglement spectrum for $a = 1, g=h=0$ as a function of $t$. The symmetry breaking mass has been set to $m = 0.0001$.
}
\label{fig_2}
\end{figure*}

\subsection{Spectrum of $\mathcal{H}_B(t)$ }
The two eigenvalues of $\mathcal{H}_+$ are given as 
\begin{equation}
	\epsilon^e_{\pm} (t) = a \pm \sqrt{g^2+(h + t)^2}-\mu 
\end{equation}
and the eigenvalues of $\mathcal{H}_-$ are given as 
\begin{equation}
	\epsilon^o_{\pm} (t) = -a \pm \sqrt{g^2+(h - t)^2}-\mu 
\end{equation}
For the sake of simplicity, we assume $a > g > h > 0$ as well as $t > 0$. Now, while tuning $t$, there can be a level crossing between two states of the different blocks,
\begin{equation}
	\epsilon^e_{-} (t) = \epsilon^o_{+} (t) \Leftrightarrow t = t_c := \frac{a \sqrt{a^2-g^2-h^2}}{\sqrt{(a-h) (a+h)}}.
\end{equation}
This crossing point is a gap closing point between the two eigenspaces of $\mathcal{J}$, and therefore the Hamiltonians $\mathcal{H}_B(t < t_c)$ and $\mathcal{H}_B(t > t_c)$ should be topologically different. We expect that this can be characterized by a topological invariant $\nu (t)$ that completely characterizes the 0-D Hamiltonian for each $t$. The invariant should not change if trivial bands are added and should be robust to deformations that do not close the gap between the occupied states. 
Due to the lack of a chiral or particle hole-symmetry, there is no natural zero of the energy, which we need to define the topological invariant.
We therefore define the zero of the energy to be at the crossing point $t_c$ and set $\mu = a-\sqrt{\left(t_c+h\right)^2+g^2}$, which enforces half-filling.

\subsection{Topological invariant and $\mathbb{Z}_2$ structure}
We here propose, with the derivation given in sec~\ref{sec:bisymmetric}, that this invariant derives not directly from the Hamiltonian, but from the matrix
\begin{equation}
\begin{split}
    \mathcal{S} (t) & = \Sigma \mathcal{H}_B(t) \\
    & = \left(
\begin{array}{cccc}
 0 & 0 & -a-g+\mu  & -h-t \\
 0 & 0 & -h-t & -a+g+\mu  \\
 a-g+\mu  & h-t & 0 & 0 \\
 h-t & a+g+\mu  & 0 & 0 \\
\end{array}
\right),
\end{split}
\end{equation}
where the $\sigma_i$ are the Pauli-matrices and $\Sigma = -\sigma_1 \otimes \sigma_0$.
The invariant is given as
\begin{equation}
	\nu(t) = \text{Pf} \left[ S (t) \right] =  \text{sign}\left[ - (h + t_c)^2 + (h + t)^2 \right],
\end{equation}
which means that $\nu(t) = -1 $ for $t < t_c$ and $\nu(t) = 1$ for $t > t_c$, while it jumps abruptly at the crossing point $t = t_c$. The invariant is a $\mathbb{Z}_2$ invariant since it can only take on the values $\pm 1$. For $t < t_c$ and $t > t_c$ there exists a gap between the lowest two energy-eigenstates and the highest two energy eigenstates [Fig.~\ref{fig_2}(a)]. \\

To prove that the topological classification is indeed a $\mathbb{Z}_2$-classification, we now double the size of the Hamiltonian 
\begin{equation}
	\mathcal{H}_D(t)  = \mathcal{H}_B(t) \otimes \sigma_3.
\end{equation}
Similarly $J_D = J_B \otimes \sigma_0$. There exist several symmetry preserving mass terms $M$, such that $\{M,\mathcal{H}_D(t)\} = [M,J_d] = [M,\mathcal{T}] = 0$, which indicates that the system becomes topologically trivial upon doubling the system. For example, $M$ can be chosen to be the matrix $M = \sigma_3 \otimes \sigma_0 \otimes \sigma_1$, which gaps out every crossing, while preserving the spatial symmetry $J$ and TRS [Fig.~\ref{fig_2}(b)].

\subsection{Entanglement spectrum}

The topological transition at $t = t_c$ is accompanied by a change in the single-particle entanglement spectrum between two spatial blocks A and B [Fig.~\ref{fig_2}(e)] of the Hamiltonian $\mathcal{H}(t)$ that are connected by varying $t$. 
The entanglement spectrum is the spectrum of the entanglement Hamiltonian $H^B_e$ which is defined through the reduced density matrix of subsystem B via~\cite{entanglement2}
\begin{equation}
	\rho_B = \frac{1}{Tr_B \left[ e^{-H^B_e} \right]} e^{-H^B_e}.
\end{equation}
For non-interacting systems, it can conveniently be obtained from the eigenvalues $\lambda_m$ of the flattened Hamiltonian $Q$ projected on subsystem $B$ as 
\begin{equation}
	Q_B (t) = \mathds{1} - P_B \left(\sum_{n\in occ}\ket{n,t}\bra{n,t}\right) P_B
\end{equation}
where $P_B$ is the projector on subsystem $B$ and $\ket{n,t}$ is an eigenstate of the Hamiltonian $\mathcal{H}(t)$.
The eigenvalues $p_i$ of the entanglement Hamiltonian $H^B_e$ can be obtained from inverting the relation 
\begin{equation}
	\lambda_n = \frac{1}{2} \tanh \left(\frac{p_n}{2} \right).
\end{equation}
In this part, we solve the problem for the analytically tractable case of $g = h = 0$.
In order to define the occupied bands for all $t$, we have to introduce an infinitesimally small symmetry breaking term $M = m \sigma_0\sigma_3$. In the limit of $m \rightarrow 0$ the projected flat band Hamiltonian is then given as
\begin{equation}
	Q_B(t)  = \left(
\begin{array}{cccc}
 \frac{1}{4} m \left(\frac{1}{\left| t-1\right| }+\frac{1}{t+1}\right) & 0 & 0 & \frac{1}{4} \left(1+\frac{1}{\text{sgn}(1-t)}\right) \\
 0 & \frac{1}{2} & 0 & 0 \\
 0 & 0 & \frac{1}{2} & 0 \\
 \frac{1}{4} \left(1+\frac{1}{\text{sgn}(1-t)}\right) & 0 & 0 & \frac{1}{4} m \left(-\frac{1}{\left| t-1\right| }-\frac{1}{t+1}\right) \\
\end{array}
\right)
\end{equation}
Expanding around $m \rightarrow 0$ again, the eigenvalues $\lambda_n$ to lowest order in $m$ are
\begin{equation}
\frac{1}{2},\frac{1}{2},
\begin{cases}
 \pm \frac{t m}{2 t^2-2}+\mathcal{O}\left(m^3\right) & t>1 \\
 \pm \left[-\frac{1}{2}-\frac{m^2}{4 \left(t^2-1\right)^2}+\mathcal{O}\left(m^3\right) \right] & t < 1.
\end{cases}
\end{equation}
Keeping only the nonsingular entanglement eigenvalues, we arrive at
\begin{equation}
p_{\pm} = 
	\begin{cases}
		\pm \left[2 \log (m)+\log \left(\frac{1}{4 \left(t^2-1\right)^2}\right)-2\right] & t < 1 \\
		\pm \frac{2 t m}{t^2-1} & t > 1 
	\end{cases}
\end{equation}
which is plotted in Fig.~\ref{fig_2}(d). The entanglement spectrum shows a discontinuous jump at $t = t_c = 1$. For $t < 1$, the spectrum is nonzero and depends on the values of $m$ and $t$, whereas it becomes quantized to zero for $t > 1$ as $m \rightarrow 0$, which is a general indication of a topological phase transition.~\cite{entanglement1,entanglement2}

\section{General theory of bisymmetric Hamiltonians} \label{sec:bisymmetric}
In this section, we introduce the general theory of real $2n \times 2n$ Hamiltonians $\mathcal{H}$ which commute with a reflection symmetry $\mathcal{J}$. For this, we choose a special basis in which the reflection symmetry takes on the form of the so-called exchange matrix
\begin{equation}
	\mathcal{J} = 
\left(
\begin{array}{cccc}
  0 & J &  \\
 J &  0 &  \\
\end{array}
\right),
\end{equation}
where $J$ is the $n \times n$ matrix with $1$'s along the anti-diagonal and $0$'s everywhere else, such that $J_{i,j} = \delta_{i,n-j+1}$.
$\mathcal{J}$ is an involution and therefore the eigenvalues are $\pm 1$. 

We begin with a few preliminary definitions. 
A matrix $X$, for which $X = JXJ$, is called \textit{perplectic} or \textit{centrosymmetric}.
Matrices $Y$ obeying $Y = J Y^T J$, where $Y^T$ is the transpose of $Y$, are called \textit{persymmetric}.
Symmetric centrosymmetric or equivalently symmetric persymmetric matrices $H = H^T$, $H = JHJ$ are called \textit{bisymmetric}.
Similarly, \textit{skew-persymmetric} matrices $V$ are defined via $V = - J V^T J$, \textit{skew-centrosymmetric} $S$ via $S = - JSJ$ and finally we call a matrix \textit{doubly-skew} if it is skew-symmetric and skew-centrosymmetric.\\
The above-mentioned types of matrices have been studied extensively and therefore we restrict ourself to only a brief review of the most relevant properties.~\cite{centro0,centro1,centro2}
The most general real perplectic $2n \times 2n$ square block matrix $X$ is given as 
\begin{equation}
	X = 
	\left(
	\begin{array}{cccc}
	  U  & JVJ    &  \\
	 V &  JUJ &  \\
	\end{array}
	\right),
\end{equation}
with $U,V \in \mathbb{R}_{2n\times 2n}$.\\
The most general real Hamiltonian which commutes with $\mathcal{J}$ therefore has to be bisymmetric, due to the additional constraint $\mathcal{H} = \mathcal{H}^T$ and is given as,
\begin{equation}\label{centrosymH}
	\mathcal{H} = 
	\left(
	\begin{array}{cccc}
	  A  & JBJ    &  \\
	  B &  JAJ &  \\
	\end{array}
	\right),
\end{equation}
where $A = A^T$ is symmetric and $B = JC^TJ$ is persymmetric. 
The Hamiltonian $\mathcal{H}$ posses $n$ $\mathcal{J}$-symmetric and $n$ $\mathcal{J}$-antisymmetric eigenvectors.

\subsection{Topological invariant}
We wish to characterize the Hamiltonian in \eqref{centrosymH} topologically by defining $N_{occ}$ occupied states ordered by energy. We define the zero of energy via a suitable shift of the chemical potential, such that $N_{occ}$ states have an energy $\varepsilon \leq 0$. In the case of a degeneracy of the highest energy state, we define the zero of energy at the point of degeneracy.

\begin{figure}[t]
 \includegraphics[width=0.8\columnwidth]{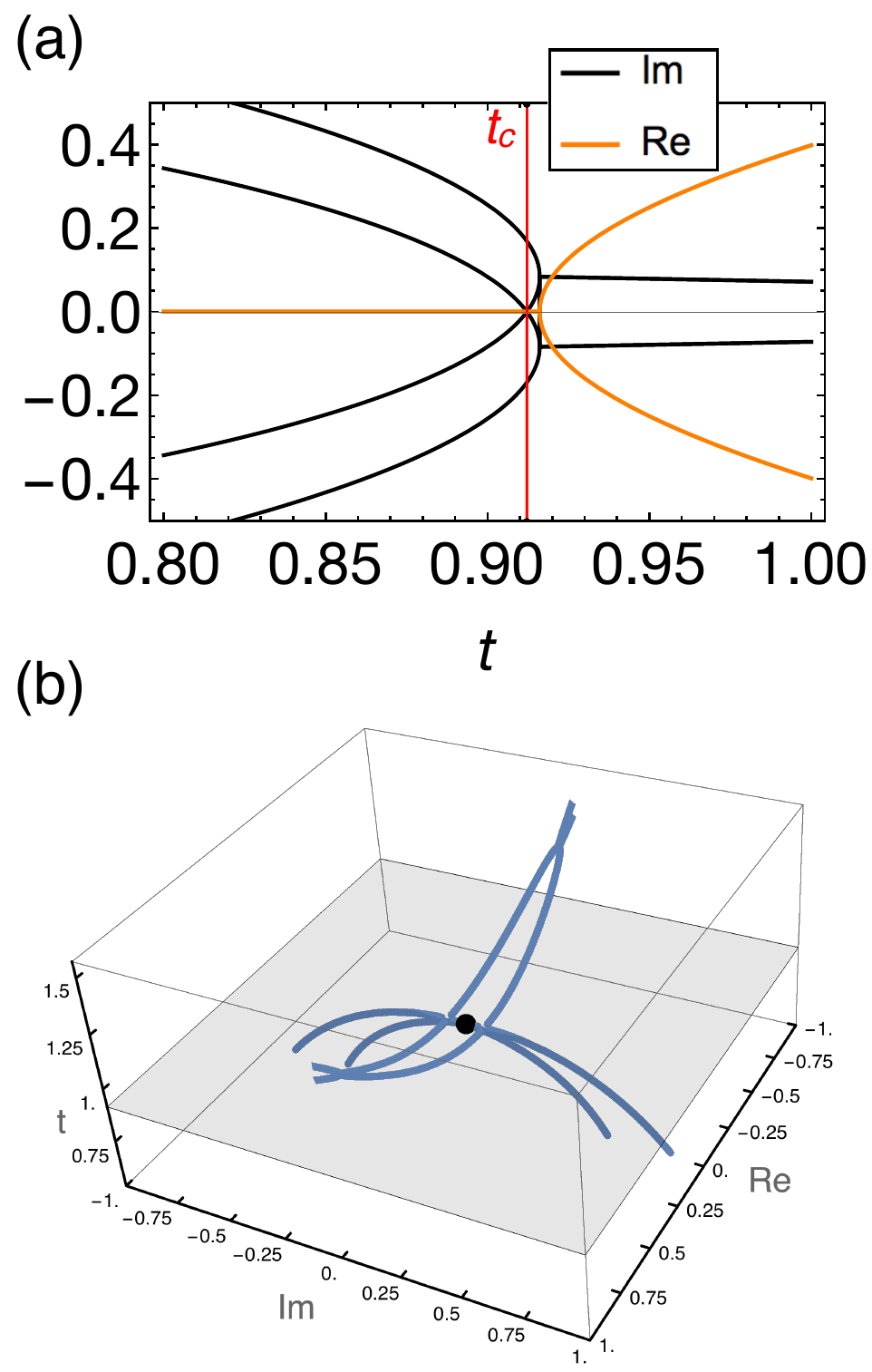}
\caption{
(a) Real and imaginary parts of the eigenvalues of $\mathcal{S}_B(t)$ as function of $t$ for $ a = 1$, $g = 0.4$ and $h = 0.2$ for the model defined in Sec.~\ref{sec:introex}. The crossing point of $\mathcal{H}_B(t)$ at $t = t_c$ corresponds to a zero mode of $\mathcal{S}_B(t)$ highlighted by a red line.
(b) Evolution of the real and imaginary parts of the eigenvalues of $\mathcal{S}_B(t)$ as a function of $t$. The $t=t_c$ plane is highlighted and the zero-eigenvalues are highlighted by a black ball.
}
\label{fig_3}
\end{figure}

Working in the basis in which $\mathcal{J}$ is diagonal, we arrive at
\begin{equation} \label{Hblock}
\mathcal{H}_B = K \mathcal{H}_B K^T  
	\left(
	\begin{array}{cccc}
	 A + JB  & 0  \\
	 0 & A - JB &  \\
	\end{array}
	\right)
\end{equation}
with  
\begin{equation}
	K = \frac{1}{\sqrt{2}}    
		\left(
	\begin{array}{cccc}
	  \mathbb{1}  & -J    &  \\
	 \mathbb{1} & J &  \\
	\end{array}
	\right)
\end{equation}
Thus the eigenstates of the different blocks of $\mathcal{H}_B$ are given by the eigenstates of the symmetric matrices $ A \pm JB$. \\
For each Hamiltonian of the form of \eqref{centrosymH} there exists a one-to-one mapping to a non-symmetric skew-centrosymmetric matrix $\mathcal{S}$ via
\begin{equation}\label{skewcentroS}
	\mathcal{S} = 
   \Sigma_3 \mathcal{H} =
	\left(
	\begin{array}{cccc}
	  A & JBJ    &  \\
	 -B & -JAJ &  \\
	\end{array}
	\right),
\end{equation}
with $\Sigma_3 = \mathbb{1}_{n\times n} \oplus-\mathbb{1}_{n\times n}$. 
The diagonalizable, non-symmetric matrix $\mathcal{S}$, by definition, possesses a chiral symmetry $\mathcal{J}\mathcal{S}\mathcal{J} = -\mathcal{S}$. 
Therefore, the eigenvalues $\lambda_i \in \mathbb{C}$ come in pairs: if $(\bs{x},\lambda)$ is an eigenpair of $\mathcal{S}$, $(\mathcal{J}\bs{x},
-\lambda)$ is an eigenpair as well. \\
Now, instead of characterizing the Hamiltonian $\mathcal{H}_B$, we choose to characterize the $\mathcal{S}$ in the diagonal basis of $\mathcal{J}$
\begin{equation} \label{Sblock}
\mathcal{S}_B = K \mathcal{S} K^T  = 
	\left(
	\begin{array}{cccc}
	  0  & A+JB    &  \\
	 A-JB & 0 &  \\
	\end{array}
	\right).
\end{equation}
This is motivated by the observation that the null-space of $\mathcal{H}_B$ is the null-space of $\mathcal{S}_B$, since
\begin{equation}
	\mathcal{S}^T_B \mathcal{S}_B = \mathcal{H}^2_B .
\end{equation}
It can be shown that only the null-space of $\mathcal{S}_B$ can be expressed in the basis of $\mathcal{H}_B$, while the non-zero eigenvectors of $\mathcal{S}_B$ are neither even nor odd under $\mathcal{J}_B$. \cite{trench2004characterization}
We now assume that there is a degeneracy between an eigenstate $\ket{+} $ of $A+JB$ and an eigenstate $\ket{-}$ of $A-JB$ at zero energy. 
It follows, that the vector $\left[\ket{+},\ket{-}\right]^T$ is a zero-mode of $\mathcal{S}_B$, since 
\begin{equation}
	\mathcal{S}_B 
	\left[
	\begin{array}{c}
	\ket{-} \\
	\ket{+} 
	\end{array} \right]
	 	=  
	\left[
	\begin{array}{c}
	(A+JB )\ket{+} \\
	(A-JB)\ket{-}
	\end{array} \right]
    =
    	\left[
	\begin{array}{c}
	0 \\
	0
	\end{array} \right]
\end{equation}
The zero modes of $\mathcal{S}_B$ thus correspond to the double degeneracies between the different blocks of $\mathcal{H}_B$ at zero energy. 
At this point, the real and imaginary parts of the eigenvalues coalesce at a so-called exceptional point [Fig~\ref{fig_3}]. Exceptional points have recently attracted interest as they are relevant for the topological classification of non-hermitian Hamiltonians in translationally invariant systems.~\cite{nonherm2,nonherm1} 
To measure this coalescence, we introduce 
\begin{equation}\label{invariant}
\nu =\sgn \pf [\mathcal{S}_B]
\end{equation}
as a topological invariant.
The pfaffian $\pf [\mathcal{S}_B] $ vanishes iff $\mathcal{S}_B$ posses a zero mode, which corresponds two zero modes of  $\mathcal{H}_B$ with opposite $\mathcal{J}$-eigenvalues. \\
The topological invariant proposed here can therefore measure if two eigenstates of $\mathcal{H}_B$ cross at zero energy as one continuously varies a parameter of the system, e.g. a hopping as discussed in Sec.~\ref{sec:introex}, for a suitable defined chemical potential.

\section{Application to chemical reactions}\label{sec:chem}
\begin{figure*}[t]
 \includegraphics[width=1.8\columnwidth]{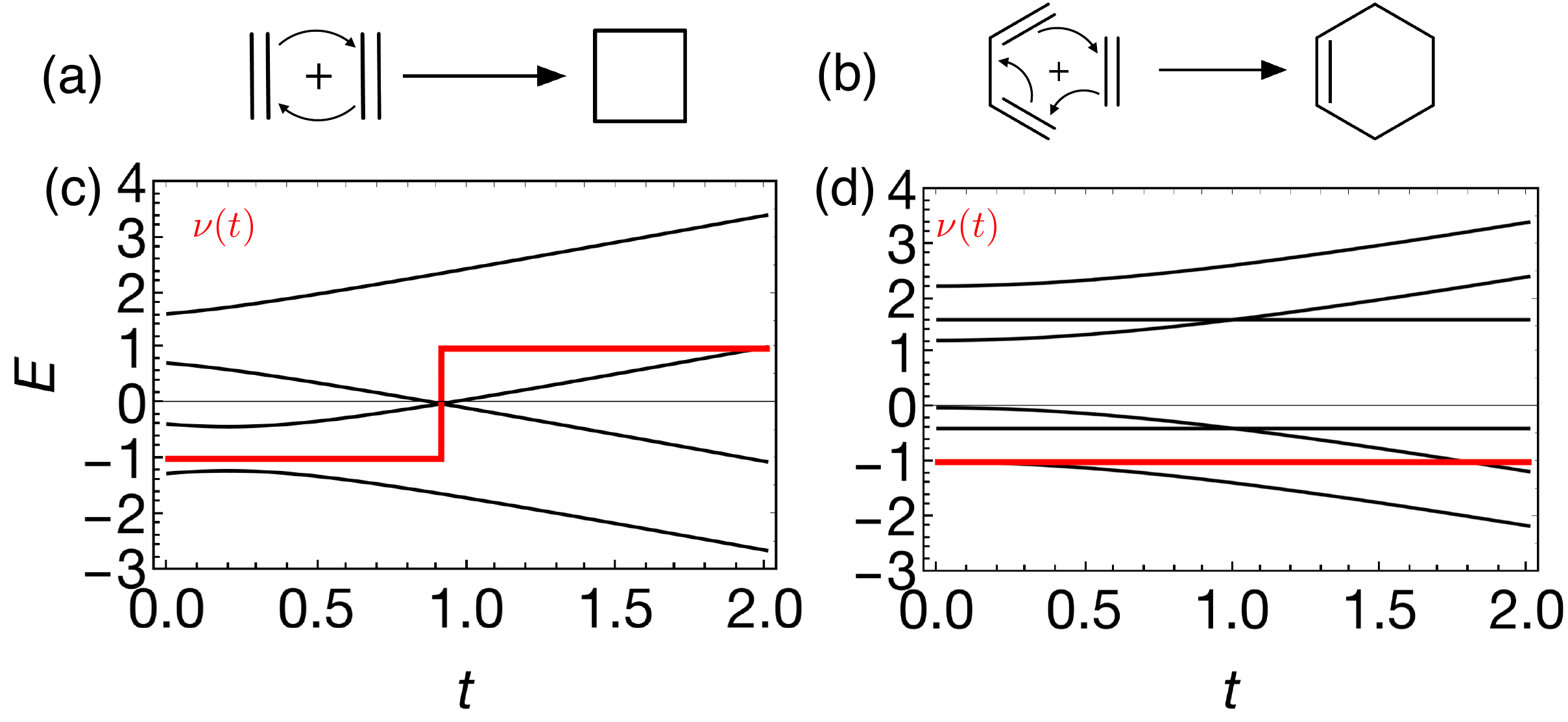}
\caption{
(a) Reaction of two ethylene ($C_2H_4$) molecules to cyclobutane ($C_4H_8$). There are two double bonds involved in the reactions, which contribute $4$ $\pi$-electrons, which is forbidden by the WHR.
(b) Reaction butadiene ($C_4H_6$) with ethylene to cyclohexene ($C_6H_{10}$).There are three double bonds involved in the reactions, which contribute $6$ $\pi$-electrons, which is allowed by the WHR.
(c)  Eigenvalues of $\mathcal{H}(t)$ and topological invariant $\nu(t)$ for the reaction depicted in (a).
(d)  Eigenvalues of $\mathcal{H}(t)$ and topological invariant $\nu(t)$ for the reaction depicted in (b).
}
\label{fig_4}
\end{figure*}

The model introduced in Sec.~\ref{sec:introex} describes the reaction of two ethylene ($C_2H_4$) molecules to cyclobutane ($C_4H_8$) in the subspace of the carbon-$p_z$ orbitals. Reactions of this type are called pericyclic reactions and their outcome can be predicted and rationalized via the Woodward-Hoffmann rules (WHR).~\cite{woodward1969conservation} 

The WHR are based on the number of $\pi$-electrons involved in a reaction. Reactions of $4n$ $\pi$-electrons, where $n \in \mathbb{N}$, are 'forbidden',  while reactions involving  $4n+2$ $\pi$-electrons are 'allowed'.
For example, in the cycloaddition of the two ethylenes there are four $\pi$-electrons involved, because each double bond contributes two $\pi$-electrons. Accordingly, the reaction does not take place under normal conditions [Fig~\ref{fig_4}(a)]. 
In contrast, the cyclodaddition of butadiene ($C_4H_6$) with ethylene to cyclohexene ($C_6H_{10}$) is allowed according to the WHR, since there are three double bonds involved, which corresponds to six $\pi$-electrons [Fig~\ref{fig_4}(b)]. This reaction takes place readily in the lab and is frequently used in organic synthesis.~\cite{nicolaou2002diels}
A common rationalization of the WHR is based on energetic arguments: 'forbidden' reactions have to overcome a large activation barrier $\Delta_R$, because there is a crossing between the occupied and unoccupied states along the reaction path [Fig~\ref{fig_4}(c)].~\cite{woodward1969conservation}
Allowed reactions on the other hand have a low reaction barrier $\Delta_R$ due to the absence of any crossings [Fig~\ref{fig_4}(d)]. 
However, this explanation in terms of energetics has two main weaknesses: \\
i) It does not take into account the strong non-adiabatic nature of the dynamics in case of a crossing, as discussed in the introduction. A transition state is not well defined in these cases and it is well known that non-adiabatic effects such as surface hopping strongly influence chemical reactions, often leading to a suppression of the reaction rate.~\cite{gabriela1994woodward}\\
ii) If the barrier height was the the only way to distinguish an 'allowed' reaction from a 'forbidden' one, there should be a crossover between 'allowed' and 'forbidden' reactions, e.g. by changing the temperature or the solvent.
This, however, is not supported by experiment. Much rather, despite valiant efforts, no 'forbidden' reaction has been reported in the literature starting from the GS of the reactants.\\
We therefore want to suggest an alternative way of understanding theses reaction rules based on topological arguments and non-adiabatic effects. The main idea has been discussed in the intruduction and we briefly review it here: If reactants and products possess different topological invariants, there has to be a crossing along the reaction path. This crossing induces very strong non-adiabatic effects, which prevent the reaction from proceeding, e.g. by ending up in an excited state instead of the ground state.
If the topological invariants of reactants and products do not differ, the reaction dynamics are adiabatic, and therefore the reactions will proceed given the right experimental conditions.
Experimentally, it has been verified that the reactions described by the Woodward-Hoffmann rules follow symmetry preserving reaction paths and therefore the topological classification based on the mirror symmetry $\mathcal{J}$ can be applied. 
We generally find that the outcome of a pericyclic chemical reaction described by the Woodward-Hoffmann rules correlates with the difference of the topological invariants of reactants and products. If reactants and products share the same value of the topological invariant $\nu(t)$ defined in the last section, they can be smoothly deformed into each other along the reaction path without a crossing between occupied and unoccupied states [Fig~\ref{fig_4}(d)], corresponding to an 'allowed' reaction.  If the topological invariant changes during the reaction, no adiabatic symmetry preserving path exists and there has to be a crossing along the reaction path  [Fig~\ref{fig_4}(c)]; in the language of the WHR, the reaction is forbidden.

\section{Summary and Conclusions}
In this paper, we have introduced a topological classification of molecules with a reflection symmetry and their chemical reactions.
In these reactions, the reflection symmetry is preserved along the reaction path and results in a $\mathbb{Z}_2$ invariant given in Eq.~\eqref{invariant} in form of a pfaffian, which is motivated by the theory of perplectic matrices. Our theory can be applied to chemical reactions that are described by the Woodward-Hoffmann rules, i.e. pericyclic reactions. We find, that Woodward-Hoffmann allowed reactions are reactions in which the topological invariant does not change along the reaction path, while the invariant of Woodward-Hoffmann forbidden reactions changes during the reaction. In light of these findings, we propose that certain chemical reactions can be described from a topological perspective, i.e. by computing a topological invariant for the reactant and product molecules. In the case where the invariant of the reactants and products is different, there has to be a gap closing point along the reaction path, if the symmetry defining the topological invariant is preserved along the path. This gap closing point has strong effects on the dynamics and time-evolution of the system and should generically lead to a suppression of the reaction rate, since there is no possibility to adiabatically move from the reactants to the products.  \\
It remains to be shown that this approach is valid for other chemical reactions rules, e.g. the Wade-Mingos rules and its extensions and if more general statements about chemical reaction rules can be made via a topological approach.~\cite{wade1971structural,mingos1984polyhedral} 

In addition it would be interesting to study the effect of electronic interactions in certain cases. Some molecular many-body states along the reaction coordinate for our toy model can be viewed as molecular analogues of Mott-Insulators, so called fragile Mott-Insulators.~\cite{moebius} It remains an open questions if the many-body nature of this states changes the topological nature of the molecular states.

\begin{acknowledgments}
The author would like to thank Raquel Queiroz, Barry Bradlyn, Jennifer Cano Andreas Schnyder for helpful discussions as well as Leslie Schoop and Martin Claassen for helpful comments on the manuscript. The
Flatiron Institute is a division of the Simons Foundation. An early part of the research has been supported by the Charlotte Elizabeth Procter Fellowship at Princeton University.
\end{acknowledgments}

\bibliography{lit}

\end{document}